\documentclass[twocolumn, showkeys,  floatfix,aps,a4paper,prd]{revtex4}
\usepackage{graphicx}

\begin{document}

\title{Massive particle production from accelerated sources in high magnetic fields}

\author{Douglas Fregolente}
\email{douglasfregolente@cfp.ufcg.edu.br}
\affiliation{Universidade Federal de Campina Grande, Centro de Forma\c c\~ao de Professores, Cajazeiras, PB,  Brazil}
\author{Alberto Saa}
\email{asaa@ime.unicamp.br}
\affiliation{
Departamento de Matem\'atica Aplicada,
UNICAMP,
  Campinas, SP, Brazil}

\begin{abstract}
Non-electromagnetic emissions from high energy particles
  in extreme environments has been studied in the literature by using several
variations of the semi-classical formalism. The detailed mechanisms behind such
emissions are of great astrophysical interest  since they can alter appreciably the associated energy loss rates. Here, we review
the role played by the source  {proper
acceleration} $a$   in the particle production process. The acceleration $a$
determines the typical scale characterizing the
  particle production and, moreover, if the massive particle production is inertially forbidden, it will be strongly suppressed for $a$ below a certain threshold.
  In particular,
  we show that,
for the case of accelerated protons in typical pulsar magnetospheres, the corresponding accelerations $a$ are far below the pion production threshold.
\end{abstract}


\maketitle

\section{Introduction}
\label{Int}

\par Processes involving particle production in the presence of
classical electromagnetic  fields $F_{ab}$ can be considered in the
context of
semi-classical approximation, in which the source particle is described by a
classical current with a prescribed trajectory
and the emitted particles are considered as fully
quantized fields. For the case of accelerated protons, the inertially
forbidden decays via the strong interaction channel like, for instance,
$p^+\stackrel{a}{\to}p^+\pi^0$ and $p \stackrel{a}{\to} n\pi^+$,
and  via the weak channel, as
 $p^+\stackrel{a}{\to}ne^+\nu$, have been studied in detail
in the semi-classical approximation. For the case of motion under  magnetic
 fields, the validity of the semi-classical approximation is warranted
if\cite{1965ARA&A...3..297G, Zharkov, Erber:1966vv},
\begin{equation}
\chi = \frac{B^{\star}}{B_{\rm cr}} \ll 1,\label{chifactor}
\end{equation}
\noindent where $B^\star = \gamma B$ is the magnitude of the
magnetic   field   measured in the
instantaneous reference frame of the proton, $\gamma$ is the usual
Lorentz factor, and $B_{\rm cr}=m^2_p/e \approx 1.5\times 10^{20}$G is a critical
value for the magnetic field, denoting the region where a full quantized
analysis is mandatory.
Provided that condition (\ref{chifactor}) is obeyed, the semi-classical
calculations are accurate and the associated emitted power depends
only on the proper acceleration of the source \cite{jackson}.
For the case of a  magnetic field $B$, for instance, the proper
acceleration of   a (relativistic)  proton is
$a\approx \gamma eB/m_p$ and, consequently, Eq.
  (\ref{chifactor}) can be rewritten as
\begin{equation}
\chi = \frac{a}{m_p} \ll 1.
\end{equation}
The parameter $\chi$ can be defined in a frame
independent way as $\chi = \sqrt{(eF^{ab}p_a)^2}/m^3_p$, where
$p_a$ stands for the proton momentum. For the case of inertially forbidden particle production, as those ones explicitly listed above, one naturally
expects a strong suppression for quasi-inertiall $(\chi\approx 0)$ protons.

We review here the
  role played by the proper acceleration $a$ of the source
  in the    massive
particle production, with emphasis  on the ranges favoring  the
massive particle emission channels.
We also discuss the possibility of observing signals of curvature
pion radiation from accelerated protons in the magnetosphere of
strongly magnetized pulsars. Unless otherwise stated,
natural units where $c = \hbar = 1$  and the spacetime signature $(+,-,-,-)$
are adopted through this work.

\section{The Semi-Classical Formalism}

\begin{figure}
\includegraphics[width=0.45\textwidth]{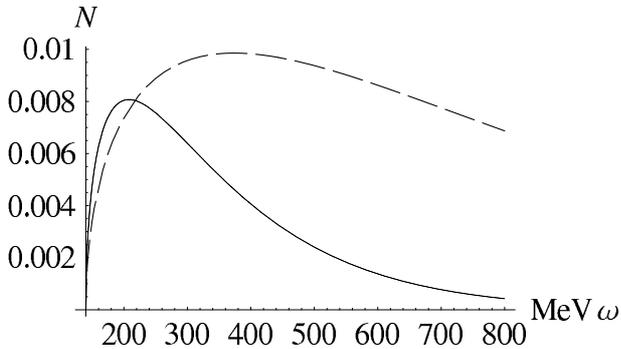}
\caption{Normalized energy distribution of the emitted scalar meson $\pi^0$
field for different values of the proper acceleration of the
proton. Full line: a = 150 MeV, dashed line: a = 500
MeV.}\label{energy}
\end{figure}

We consider here   the following kind of decay process
\begin{equation}
p_1 \stackrel{a}{\to} p_2 +\sum_i q_i \label{general},
\end{equation}
where $p_{1}$ and $p_{2}$ are the source particles, described by a  classical current  corresponding to
the states of   a two level system following a prescribed classical trajectory
with proper acceleration $a$, and
$q_i$ are the emitted particles, described by quantized fields. (For further details,
see \cite{2001PhRvD..63a4010V,Fregolente:2006xj}, for instance.)
We assume that the respective masses $m_{q_i}$ of the emitted
particles $q_i$ obeys $m_{q_i} < m_{p_{1,2}}$, where $m_{p_{1}}$ and $m_{p_{2}}$
stands for, respectively, the masses of the source particles  $p_1$ and $p_2$.
If $m_{p_{1}} < m_{p_{2}} +\sum_i m_{q_i} $, the process is known to be
forbidden for inertial trajectories
($a=0$). It is intuitive to expect that,  if the source is supposed to follow a prescribed trajectory,
  the  momentum $\mathbf{k}_{i}$ of the
emitted particle ${q_i}$, measured in the rest frame
of the source,  must be constrained to $|\mathbf{k}_{i}|
\ll m_{1,2}$\cite{2001PhRvD..63a4010V, Fregolente:2006xj}.
Moreover,
the mean energy $\tilde{\omega}_{i}$ of the emitted particles
(also measured in the proton's reference frame) is of the order of
the source proper acceleration\cite{2001PhRvD..63a4010V}, {\em i.e.},
\begin{equation}
\tilde{\omega}_{i}\sim a. \label{threshold}
\end{equation}
Then, from
the condition $m_{q_i} \ll m_{1,2}$ and $\tilde{\omega}_{i}^2 =
|\mathbf{k}_{i }|^2+m_{q_1}^2$ one has $a \ll m_{1,2}$. In order to
clarify these points,
let us consider the explicit example of the emission of neutral pions
by uniformly accelerated protons,
\begin{equation}
p^+ \stackrel{a}{\to} p^+   \pi^0 \label{scalarmeson1}.
\end{equation}
Without loss of generality, let us assume a
  uniformly accelerated trajectory along the $z$
direction with worldline   given by
\begin{equation}
x^\mu = a^{-1}\left(\sinh(a\tau),0,0,\cosh(a\tau)\right).
\end{equation}
\noindent Following the same procedure employed in the Refs.
\cite{2001PhRvD..63a4010V, Fregolente:2006xj}, one obtains the
differential decay rate as function of the energy of the emitted pion
\begin{equation}
\frac{d\Gamma}{d\tilde{\omega}}= \frac{G_{\rm
eff}^2}{2\pi^2a}\sqrt{\tilde{\omega}^2-m_\pi^2}K_0(\tilde{\omega}/a),
\end{equation}
\noindent where $K_0$ is the modified Bessel function of order $0$,
$m_\pi\approx 140$ MeV is the $\pi^0$ mass,
and $G_{\rm eff}$ is the effective coupling constant. The total emitted power
 is
\begin{equation}
W   = \frac{G^2_{\rm eff}m^3_\pi}{\pi^{5/2}a}G_{13}^{30} \left(
\frac{m^2_\pi}{4a^2} \left|
\begin{array}{l}
\;\;\;0\\
-{1}/{2}\;,0,\;0
\end{array}
\right. \right) \;,
\end{equation}
\noindent where $G_{mn}^{pq}$ stands for the Meijer G-function\cite{grad}. The
corresponding
  normalized energy distribution,
\begin{equation}
N=\frac{1}{\Gamma}\frac{d\Gamma}{d\tilde{\omega}}
\end{equation}
\noindent is plotted in Fig.(\ref{energy}),
from which it is clear that
  mean energy of the
emitted particles are $\tilde{\omega}\approx a$.
From (\ref{threshold}) and
Fig.(\ref{energy}), one can argue that, if the   acceleration  is such that
\begin{equation}
a \geq \sum_i m_{q_1}\label{scalarmeson},
\end{equation}
\noindent the corresponding  process (\ref{general})
becomes energetically favored.
In fact,   more precise estimates reveal  that the threshold is
  $a \ge \Delta m +\sum_i m_{q_1}$, where $\Delta m =
m_2-m_1$.  Hence, the neutral pion production (\ref{scalarmeson1})
is expected do be favored for  $a \geq m_\pi
\approx 140$ MeV.
For the physical relevant case of protons in circular motion, one has also
exactly the same
  threshold (\ref{scalarmeson}). For such case, the formulas for the neutral pion production (\ref{scalarmeson1}) can be obtained from the previously
one calculated in \cite{Fregolente:2006xj} for scalar emissions
\begin{equation}
W_s \approx \frac{G_{\rm
eff}^{(s)2}a^2}{12\pi}.\label{powerscalar}
\end{equation}
We notice that, in order to obtain that the total power emission
 (\ref{powerscalar}) from
Eq. (4.7) of \cite{Fregolente:2006xj}, one must perform the following
modifications on the    particle states: $|\pi^+\rangle\mapsto
|\pi^0\rangle$ and $|n\rangle\mapsto |p\rangle$. These substitutions
ensure the charge conservation and  that the initial and final
charged nucleon states are coupled in the same way to the magnetic field. In
(\ref{powerscalar}), $G_{\rm
eff}^{(s)}$ is related to  the strong coupling constant $g$
by
\begin{equation}
G_{\rm eff}^{(s)2} =  \frac{g^2}{4\pi} \approx 14.
\end{equation}
The large difference between the strong and the electromagnetic
coupling constants will imply a large difference also in the
respective emitted powers.
The comparison between
(\ref{powerscalar}) and the Larmor formula for the usual synchrotron radiation
shows that $W_s/W_\gamma = g^{  2}/4\pi\alpha \gg 1$, where
$\alpha\approx 1/137$ is the electromagnetic coupling constant. Hence,
 if the
proper acceleration of a proton reach the   value  $a \approx
m_{\pi}$, the channel (\ref{scalarmeson1}) is expect to dominate over
the usual synchrotron radiation emission.

 It must be emphasized that the formalism presented here cannot be applied when the charge currents corresponding to    initial and   final
states   are different. Neutral and charged particles
follow distinct trajectories in the presence of an electromagnetic
field. The related channel $p^+\to n\pi^+$ for example, requires a
different semi-classical formalism. The formalism and the respective
calculations can be found in \cite{2008JPhG...35b5201H} (see also
\cite{Fregolente:2007qw}).

\section{Applications}

Pulsar magnetospheres  offer many interesting possibilities to test these results involving high energy accelerated particles\cite{1995PhLB..351..261B, 2008JCAP...08..025H,1538-4357-525-2-L117}. Let us consider, for this
 purpose, the simplest
    polar cap models of pulsars
\cite{Ruderman:1975ju, Harding:2004hj,Vietri:1109401}. For such models, a proton
interacts with the dipolar magnetic field of the pulsar and is
accelerated by an electric field parallel to the magnetic field
lines. The  transverse to the magnetic field component
$p_\perp$  of the proton momentum
vanish rapidly due to the usual synchrotron emission. The protons
are accelerated along the magnetic field lines by the electric potential drop of the pulsar, producing photons by curvature radiation.

  A particle following a circular trajectory of radius $R_{\rm c}$ has proper centripetal  acceleration given by
$a \approx \gamma^2/R_{\rm c}$. The curvature radius $R_{\rm c}$ of
the magnetic field line of a pulsar is related to the radius $r_{\rm
s}$ of the star and to the light cone radius,
 $R_{\rm L}=P/2\pi$ through $R_{\rm c} = 4/3 (r_{\rm s}R_{\rm L})^{1/2}$, where $P$ is the star period. Then, one has
\begin{equation}
a \approx \frac{\eta^2e^2\phi^2}{m^2R_{\rm c}}.\label{acceleration}
\end{equation}
\noindent The  $\eta$ factor accounts for the efficiency of the
polar cap acceleration mechanism \cite{arons, meszaros} $\gamma =
\epsilon/m = \eta e\phi/m$. Also,
\begin{equation}
\phi = \left(\frac{2\pi^2Br^3_{\rm s}}{P^2}\right) \approx 6.6\times
10^{18}B_{12}P^{-2}_{-3}\left(\frac{r_{\rm s}}{10^4{\rm m}}\right)^3
{\rm V},
\end{equation}
\noindent is the maximum available potential drop near the surface
of the star \cite{2008JCAP...08..025H,0004-637X-556-2-987}. Taking
for a typical young pulsar $r_{\rm s} \sim 10^4$ m, one has
from (\ref{acceleration}) and (\ref{scalarmeson}), the following
condition on the star surface magnetic field
\begin{equation}
B_{12} \geq 0.65\times P_{-3}^{9/4},\label{final}
\end{equation}
\noindent for the occurrence of pion  emission   with   non negligible
intensity. It is easily verified that even for $\eta =  1$, the
condition (\ref{final}) is far above of the known pulsars of the ATNF   Catalogue\cite{ATNF}, see Fig. 2. In fact,   a systematic search in the ATNF pulsar catalogue for strongly magnetized stars that could accelerate relativistic protons up to the curvature pion production threshold
reveals
that the
  best candidate is the 16 ms pulsar J0537-6910, but the  corresponding characteristic parameter $\chi=a/m_p=10^{-6}$ is   too
small     to give origin to  observable signals \cite{FS} .
\begin{figure}
\includegraphics[width=0.45\textwidth]{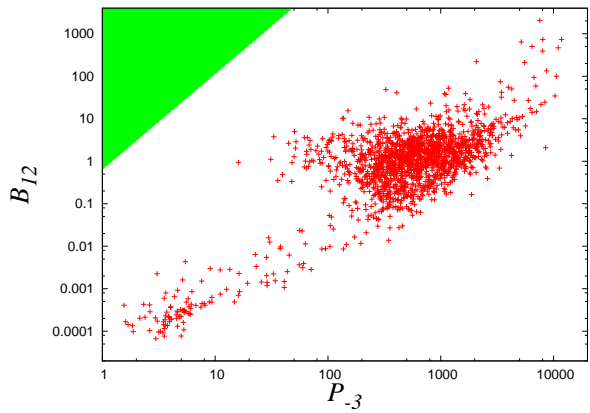}\label{bp}
\caption{
The values $(P_{-3},B_{12})$ for the 1700 objects in the ATNF Pulsar Catalogue\cite{ATNF} with known period $P = P_{-3}\times 10^{-3}\,$s and surface magnetic field $B=B_{12}\times 10^{12}\,$G. The shadowed region
 corresponds to the condition (3.3) . All the pulsars
in the ATNF catalogue  are far below the pion production threshold.}
\end{figure}

\section{Conclusions}

We have briefly reviewed  the role played by the source
proper acceleration
in the semi-classical approximation for the process of massive
particle emission. Unfortunately, we have
  from Eq.(\ref{final}) and Fig. 2 that the necessary threshold for pion production by protons in circular motion is far above the typical
  acceleration experimented by protons in pulsar magnetospheres.
  Nevertheless, it would be also interesting  to investigate, in this context,  other radiative processes
  like photopion production and the inverse Compton scattering.

\section{Acknowledgements} The authors are grateful to A. T. Dias,
 C. O. Escobar,   G. E. A. Matsas, R.
Opher,    S. Razzaque, and V. Pleitez for enlightening and helpful discussions.

\end{document}